\documentclass[12pt]{article}
\usepackage{amsfonts}
\usepackage{amssymb}
\usepackage{mathrsfs}
\usepackage{srcltx}
\usepackage{amsmath}
\usepackage{amsthm}
\usepackage{amstext}
\usepackage{amsopn}
\usepackage{graphicx}

\newtheorem{theorem}{Theorem}[section]
\newtheorem{lemma}[theorem]{Lemma}

\newtheorem{corollary}[theorem]{Corollary}

\theoremstyle{definition}
\newtheorem{definition}[theorem]{Definition}

\theoremstyle{remark}

\numberwithin{equation}{section}

\newcommand{\ba}{\begin{array}}
\newcommand{\ea}{\end{array}}

\begin{document}

\title{ \bf\large{\ New Periodic Solutions of Singular Hamiltonian Systems  with  Fixed Energies}}
\author{Fengying Li\textsuperscript{1}\footnote{Corresponding Author, Email: lify0308@163.com}\ \ Qingqing Hua\textsuperscript{2}\ \ Shiqing Zhang\textsuperscript{3}
 \\
{\small \textsuperscript{1} School of Economic and Mathematics, Southwestern University of Finance and Economics,\hfill{\ }}\\
\ \ {\small Chengdu, Sichuan, 611130, P.R.China.\hfill{\ }}\\
{\small \textsuperscript{2,3} Department of Mathematics, Sichuan University, \hfill{\ }}\\
\ \ {\small Chengdu, Sichuan, 610068, P.R. China.\hfill {\ }} }

\date{}
\maketitle

\begin{abstract}
{By using the variational minimizing method with a special constraint and
 the direct variational minimizing method without constraint, we
  study second order Hamiltonian systems with a singular potential $V\in C^2(R^n\backslash O,R)$
 and $V\in C^1(R^2\backslash O,R)$ which may have an unbounded potential well, and prove the existence of non-trivial periodic solutions with a prescribed energy. Our results can be
regarded as some complements of the well-known Theorems of
 Benci-Gluck-Ziller-Hayashi and Ambrosetti-Coti Zelati and so on.}\\


\end{abstract}

\section*{1. Introduction}
\setcounter{section}{1} \setcounter{equation}{0}

Seifert([33]) in 1948 and Rabinowitz([29,30]) in 1978 and 1979 studied classical second order Hamiltonian systems without singularity, based on their work, Benci ([8,9])and
Gluck-Ziller([19]) and Hayashi([23]) used Jacobi metric and very
complicated geodesic methods and algebraic topology to study the
periodic solutions with a fixed energy of the following system:
\begin{equation}
\ddot{q}+V^{\prime}(q)=O,\label{1.1}
\end{equation}
\begin{equation}
\frac{1}{2}|\dot{q}|^2+V(q)=h.\label{1.2}
\end{equation}
They proved  a very general theorem:
\begin{theorem}\label{1.1} Suppose $V\in C^2(R^n,R)$ ,if
\begin{equation*}
\{x\in R^n|V(x)\leq h\}
\end{equation*}
 is bounded and non-empty, then the (1.1)-(1.2) has a  periodic solution with energy h.

  Furthermore, if
 \begin{equation*}
 V^{\prime}(x)\not=O,\ \ \ \ \forall x\in\{x\in R^n|V(x)=h\},
 \end{equation*}
 then the (1.1)-(1.2) has a nonconstant periodic solution with energy h.
 \end{theorem}
For the existence of multiple periodic solutions for (1.1)-(1.2)
 with compact energy surfaces, we can refer Groessen([22]) and Long[24]
 and the references there.

 In the 1987 paper of  Ambrosetti-Coti Zelati[1], Clark-Ekeland's dual action principle, Ambrosetti-Rabinowitz's mountain pass theorem etc. were used to study the existence of $T$-periodic solutions of the second-order equation
  \begin{equation*}
 -\ddot x=\nabla U(x),
  \end{equation*}
 where
  \begin{equation*}
 U=V\in C^2(\Omega;\bold R)
  \end{equation*}
 is such that
  \begin{equation*}
 U(x)\to\infty,x\to \Gamma=\partial\Omega;
  \end{equation*}
where $\Omega\subset\bold R^n$
  is a bounded and convex domain, they got the following result:
\begin{theorem}\label{1.2} Suppose that
\begin{enumerate}
  \item[(i).] $U(O)=0=\min U$ ;
  \item[(ii).] $U(x)\le \theta(x,\nabla U(x))$ for some $\theta\in (0,\tfrac 12)$ and for all $x$ near $\Gamma$ (superquadraticity near $\Gamma$);
  \item[(iii).] $(U''(x)y,y) \ge k|y|^2$ for some $k>0$ and for all $(x,y)\in\Omega\times \bold R^N$.

Let $\omega_N$ be the greatest eigenvalue  of $U'' (0)$ and $T_0=(2/\omega_N)^{1/2}$. Then  $-\ddot x=\nabla U(x)$ has for each $T\in (0,T_0)$ a periodic solution with minimal period $T$.
\end{enumerate}
\end{theorem}
 For $C^r$  systems, a natural interesting problem is if
 \begin{equation*}
\{x\in R^n|V(x)\leq h\}
\end{equation*}
is unbounded, can we get nonconstant periodic solution for the system
$(1.1)-(1.2)?$

    In 1987, D.Offin [27] firstly generalized Theorem 1.1 to some
    non-compact cases under $V\in C^3(R^n,R)$ and complicate geometrical
    assumptions on potential wells, but it seems to be difficult to verify for concrete potentials under the geometrical conditions.

    In 1988, Rabinowitz [31] studied multiple periodic solutions for
    classical Hamiltonian systems with potential $V\in C^1(R\times
    R^n,R)$, where $V(q_1,...,q_n;t)$ is $T_i$-periodic in positions $q_i\in R$ and is T-periodic in $t$.

In 1990, using Clark-Ekeland's dual variational principle and
Ambrosetti-Rabinowitz's Mountain Pass Lemma, Coti
Zelati-Ekeland-Lions [14] studied Hamiltonian systems with convex
potential wells, they got the following result:

\begin{theorem}\label{1.3}
 Let $\Omega$ be a convex open subset of $R^n$
containing the origin $O$. Let $ V \in C^2(\Omega,R)$ be such that
\begin{enumerate}
 \item[$(V1)$.] $V(q)\geq V(O)=0,\forall q\in\Omega$.
 \item[$(V2)$.] $\forall q\neq O,V''(q)>0$.
 \item[$(V3)$.] $\exists \omega>0$, s.t. $V(q)\leq \frac{\omega}{2}\|q\|^2,\forall \|q\|<\epsilon$.
 \item[$(V4)$.] $V''(q)^{-1}\rightarrow 0,\|q\|\rightarrow 0,$

 or
 \item[$(V4)'$.] $V''(q)^{-1}\rightarrow 0, q\rightarrow\partial\Omega.$

Then, for every $T<\frac{2\pi}{\sqrt{\omega}}$, (1.1) has a
solution with minimal period $T.$
\end{enumerate}
\end{theorem}
 In Theorems 1.2 and 1.3, the authors assumed the convex conditions
 for potentials and potential wells so that they can apply Clark-Ekeland's
 dual variational principle, we notice that Theorems 1.1-1.3 essentially made the following
 assumption:
 \begin{equation*}
 V(x)\to\infty,x\to \Gamma=\partial\Omega.
 \end{equation*}
 So that all the potential wells are bounded.

 For singular Hamiltonian systems with a fixed energy $h\in R$,
 Ambrosetti-Coti Zelatiin [3,5]
used Ljusternik-Schnirelmann theory on a $C^1$ manifold to get the following Theorem:
\begin{theorem}\label{1.4}(Ambrosetti-Coti Zelati[3])\ \ Suppose $V\in
C^2(R^n\backslash \{O\},R)$ satisfies $V(q)\rightarrow -\infty,q\rightarrow 0$ and
\begin{enumerate}
 \item[$(A1)$.] $3V^{\prime}(u)\cdot u+(V''(u)u,u)\not=0$;
 \item[$(A2)$.] $V^{\prime}(u)\cdot u>0$;
 \item[$(A3)$.] $\exists \alpha >2$, s.t. $V^{\prime}(u)\cdot u\leq-\alpha V(u)$;
 \item[$(A4)$.] $\exists \beta >2, r>0$, s.t. $V^{\prime}(u)\cdot u\geq-\beta V(u), 0<|u|<r$;
 \item[$(A5)$.] $V(u)+\frac{1}{2}V'(u)u\leq 0.$

Then (1.1)-(1.2) has at least one non-constant periodic
solution.
\end{enumerate}
\end{theorem}
After Ambrosetti-Coti Zelati, a lot of mathematicians studied
singular Hamiltonian systems, here we only mention of a related recent
paper of Carminati-Sere-Tanaka[11], they used complex variational
 and geometrical and topological methods to generalize Pisani's
 results ([28]), they got
\begin{theorem}\label{1.5}
Suppose $h>0,L_0>0$ and $V\in C^{\infty}(R^n\backslash \{O\},R)$
satisfies $V(q)\rightarrow -\infty,q\rightarrow 0$ and
\begin{enumerate}
 \item[$(B1)$.] $V(q)\leq 0;$
 \item[$(B2)$.] $V(q)+\frac{1}{2}V'(q)q\leq h,\forall|q|\geq e^{L_0};$
 \item[$(B3)$.] $V(q)+\frac{1}{2}V'(q)q\geq h,\forall|q|\leq e^{-L_0};$
 \item[$(A4)$.] $\exists \beta>2, r>0$, s.t. $V^{\prime}(q)\cdot q\geq-\beta V(q), 0<|q|<r$;

Then $(1.1)-(1.2)$ has at least one periodic solution with the given energy h and whose action is at most $2\pi r_0$ with
\begin{equation*}
 r_0=\max\{[2(h-V(q))]^{\frac{1}{2}};|q|=1\}.
 \end{equation*}
 \end{enumerate}
 \end{theorem}
\begin{theorem}\label{1.6}
Suppose $h>0,\rho_0>0$ and  $V\in C^{\infty}(R^n\backslash
\{O\},R)$ satisfies $V(q)\rightarrow -\infty,q\rightarrow 0$ and $(B1),(A4)$ and
\begin{enumerate}
 \item[$(B2')$.] $\lim_{|q|\rightarrow+\infty}V'(q)=O;$
 \item[$(B3')$.] $V(q)+\frac{1}{2}V'(q)q\geq h,\forall |q|\leq\rho_0;$
\end{enumerate}
Then $(1.1)-(1.2)$ has at least one periodic solution with the given energy h and whose action is at
 most $2\pi r_0.$
\end{theorem}
By using the variational minimizing method with a special constraint, we obtain the following result:
 \begin{theorem}\label{1.7} Suppose $V\in C^2(R^n\backslash \{O\},R)$  and $V(q)\rightarrow -\infty,q\rightarrow 0$
 and satisfies $(A1)-(A3)$ and
\begin{enumerate}
 \item[$(A4)'$.] $\exists \beta>2$, s.t. $V^{\prime}(q)\cdot q\geq-\beta V(q), 0<|q|<+\infty$;
 \item[$(A5)'$.] $V(-q)=V(q),\forall q\not=O$.

Then for any $h>0$,$(1.1)-(1.2)$ has at least one
 non-constant periodic solution with the given energy h.
 \end{enumerate}
\end{theorem}
Using the direct variational minimizing method, we get the following Theorem:\\

\begin{theorem}\label{1.8} Suppose $V\in C^1(R^2\backslash \{O\},R)$ and $V(q)\rightarrow -\infty,q\rightarrow 0$
and satisfies
\begin{enumerate}
 \item[$(B1)'$.] $V(q)<h,\forall q\not=O$;
 \item[$(P1)'$.] $V'(u)\rightarrow O,\|u\|\rightarrow +\infty$;
 \item[$(A3)'$.] $\exists \alpha>2,\mu_2>0$, s.t. $V^{\prime}(u)\cdot u\leq-\alpha V(u)+\mu_2$;
 \item[$(A4)$.] $\exists \beta>2, r>0$, s.t. $V^{\prime}(u)\cdot u\geq-\beta V(u), 0<|u|<r$.

Then for any $h>\frac{\mu_2}{\alpha}$,$(1.1)-(1.2)$ has at least one
 non-constant periodic solution with the given energy h.
 \end{enumerate}
\end{theorem}
\begin{corollary}
 Suppose $\alpha=\beta> 2$ and
$$V(x)=-|x|^{-\alpha}$$
Then for any $h>0$, $(1.1)-(1.2)$ has at least one
non-constant periodic solution with the given energy h.
\end{corollary}
{\bf Remark:} In our Theorem 1.8, the assumption on regularity for
potential $V$ is weaker than Theorems 1.1-1.6. Comparing Theorem
1.5 with Theorem 1.8, our $(B1)'$ is also weaker than $(B1)$, and $(A3)'$
is also different from $(B2)-(B3)$ and $(B3')$.

\section{ A Few  Lemmas}
\setcounter{section}{2} \setcounter{equation}{0}

 Let
 \begin{equation*}
 H^1=W^{1,2}(R/Z,R^n)=\{u:R\rightarrow R^n,u\in L^2,\dot{u}\in L^2,u(t+1)=u(t)\}
 \end{equation*}
 Then the standard $H^1$ norm is equivalent to
 \begin{equation*}
 \|u\|=\|u\|_{H^1}=\left(\int^1_0|\dot{u}|^2dt\right)^{1/2}+|u(0)|.
 \end{equation*}
 Let
 \begin{equation*}
 \Lambda=\{u\in H^1|u(t)\neq O,\forall t\}.
 \end{equation*}
 \begin{lemma}\label{l2.1}([3])\ \ Let
\begin{equation*}
F=\{u\in H^1|\int_0^1(V(u)+\frac{1}{2}V^{\prime}(u)u)dt=h\}.
\end{equation*}
If $(A1)$ holds, then $F$ is a $C^1$ manifold with codimension 1 in
$H^1.$ Let
\begin{equation*}
f(u)=\frac{1}{4}\int^1_0|\dot{u}|^2dt\int^1_0V^{\prime}(u)udt
\end{equation*}
and $\widetilde{u}\in F$ be such that
$f^{\prime}(\widetilde{u})=O$
 and $f(\widetilde{u})>0$. Set
 \begin{equation*}
\frac{1}{T^2}=\frac{\int^1_0V^{\prime}(\widetilde{u})\widetilde{u}dt}{\int^1_0|\dot{\widetilde{u}}|^2dt}
\end{equation*}
If $(A2)$ holds, then $\widetilde{q}(t)=\widetilde{u}(t/T)$ is a
non-constant $T$-periodic solution for (1.1)-(1.2).
 Moreover, if $(A2)$ holds, then $f(u)\geq 0$ on $F$ and
 $f(u)=0,u\in F$ if and only if $u$ is constant.
\end{lemma}
\begin{lemma}\label{l2.2}([3,22]) Let
$f(u)=\frac{1}{2}\int^1_0|\dot{u}|^2dt\int^1_0(h-V(u))dt$ and
$\widetilde{u}\in\Lambda$ be such that
$f^{\prime}(\widetilde{u})=O$
 and $f(\widetilde{u})>0$. Set
\begin{equation*}
\frac{1}{T^2}=\frac{\int^1_0(h-V(\widetilde{u}))dt}{\frac{1}{2}\int^1_0|\dot{\widetilde{u}}|^2dt}\label{2.1}
\end{equation*}
Then $\widetilde{q}(t)=\widetilde{u}(t/T)$ is a non-constant
$T$-periodic solution for (1.1)-(1.2). Furthermore, if
$V(x)<h,\forall x\not=O$, then $f(u)\geq 0$ on $\Lambda$ and
 $f(u)=0,u\in \Lambda$ if and only if $u$ is a nonzero constant.
\end{lemma}
\begin{lemma}\label{l2.3}
(Sobolev-Rellich-Kondrachov[26],[41])
\begin{equation*}
W^{1,2}(R/Z,R^n)\subset C(R/Z,R^n)
\end{equation*}
and the imbedding is compact.
\end{lemma}
\begin{lemma}([26,41])\label{l2.4} Let $q\in W^{1,2}(R/TZ,R^n)$.
\begin{enumerate}
\item[(1).] If $q(0)=q(T)=O$, then we have Friedrics--Poincar\'{e}
inequality:
\begin{equation*}
\int^T_0|\dot{q}(t)|^2dt\geq\left(\frac{\pi}{T}\right)^2\int^T_0|q(t)|^2dt.
\end{equation*}
\item[(2).] If $\int_0^Tq(t)dt=0$,then we have Wirtinger's inequality:
\begin{equation*}
\int^T_0|\dot{q}(t)|^2dt\geq\left(\frac{2\pi}{T}\right)^2\int^T_0|q(t)|^2dt
\end{equation*}
and Sobolev's inequality:
\begin{equation*}
\int^T_0|\dot{q}(t)|^2dt\geq\frac{12}{T}|q(t)|^2_{\infty}.
\end{equation*}
\end{enumerate}
\end{lemma}
\begin{lemma}\label{l2.5}(Eberlein-Shmulyan [39])\ \ A Banach space $X$ is
reflexive if and only if any bounded sequence in $X$ has a weakly
convergent subsequence.
\end{lemma}

\begin{definition}\label{d2.1}(Tonelli ,[26])\ \ Let $X$ is a Banach space,
$f:X\rightarrow R$.
\begin{enumerate}
\item[(i).] If for any $\{x_n\}\subset X$ strongly converges to
$x_0$:$x_n\rightarrow x_0$, we have
\begin{equation*}
 \liminf f(x_n)\geq f(x_0),
\end{equation*}
then we call $f(x)$ is lower  semi-continuous at $x_0$.
\item[(ii).] If for any $\{x_n\}\subset X$  weakly converges to
$x_0$:$x_n\rightharpoonup x_0$, we have
\begin{equation*}
\liminf f(x_n)\geq f(x_0),
\end{equation*}
then we call $f(x)$ is weakly lower  semi-continuous at $x_0$.
\end{enumerate}
\end{definition}
Using the famous Ekeland's variational principle,Ekeland proved
\begin{lemma}\label{l2.6}(Ekeland[16]) Let $X$ be a  Banach
space,$F\subset X$ be a closed (weakly closed) subset, let $\delta(x_1,x_2)$ be the geodesic distance between two points $x_1$ and $x_2$ in $X$, $\delta(x,F)$ be the geodesic distance between $x$ and the set $F$.
 Suppose that $\Phi $ defined on $X$ is Gateaux-differentiable and  lower
 semi-continuous (or weakly lower semi-continuous)
  and assume $\Phi|_F$ restricted on $F$ is bounded from
 below. Then there is a sequence $\{x_n\}\subset F$ such that
\begin{equation*}
\begin{array}{l}
 \delta(x_n,F)\rightarrow 0,\\
 \Phi(x_n)\rightarrow\inf_{F}\Phi,\\
 (1+||x_n||)\|\Phi|_F^{'}(x_n)\|\rightarrow 0.\\
\end{array}
\end{equation*}
\end{lemma}
\begin{definition}\label{d2.2}([16,18])
Let $X$ be a  Banach
space, $F\subset X$ be a closed  subset.
 Suppose that $\Phi $ defined on $X$ is Gateaux-differentiable,
 if sequence $\{x_n\}\subset F$ such that
\begin{equation*}
\begin{array}{l}
\delta(x_n,F)\rightarrow 0,\\
\Phi(x_n)\rightarrow c,\\
(1+||x_n||)\|\Phi|_F^{'}(x_n)\|\rightarrow 0,
\end{array}
\end{equation*}
then $\{x_n\}$ has a strongly convergent subsequence.

Then we call $f$ satisfies $(CPS)_{c,F}$ condition at the level $c$
for the closed subset $F\subset X$.
\end{definition}
We notice that if $F=X$, then the above condition is the classical Cerami-Palais-Smale condition[13].

We can give a weaker condition than $(CPS)_{c,F}$ condition:
\begin{definition}\label{d2.3} Let $X$ be a  Banach space, $F\subset X$ be
a weakly closed subset.
 Suppose that $\Phi $ defined on $X$ is Gateaux-differentiable,
 if sequence $\{x_n\}\subset F$ such that
\begin{equation*}
\begin{array}{l}
\delta(x_n,F)\rightarrow 0,\\
\Phi(x_n)\rightarrow c,\\
\|\Phi|_F^{'}(x_n)\|\rightarrow 0,
\end{array}
\end{equation*}
then $\{x_n\}$ has a weakly convergent subsequence.

Then we call $f$ satisfies $(WCPS)_{c,F}$ condition.
\end{definition}
\begin{lemma}\label{l2.7}(Gordon [20]) Let $V$ satisfies so called
Gordon's Strong Force condition:

There exists a neighborhood $\mathcal{N}$ of O
and a function $U\in C^1(\Omega ,\mathbb{R})$ such that:
\begin{enumerate}
 \item[(i).] $\lim\limits_{s\rightarrow 0}U(x)=-\infty$;
 \item[(ii).] $-V(x)\geq |U^{\prime}(x)|^2$ for every
$x\in\mathcal{N}-\{O\}$ .

Let
\begin{equation*}
\partial\Lambda=\{u\in H^1=W^{1,2}(R/Z,R^n),\ \ \exists t_0 , u(t_0)=O\}.
\end{equation*}
Then we have
\begin{equation*}
\int^1_0V(u)dt\rightarrow -\infty,\forall u_n\rightharpoonup u\in
\partial\Lambda.
\end{equation*}
Let
\begin{equation*}
\partial\Lambda=\{u\in
H^1=W^{1,2}(R/Z,R^n),\exists t_0 , u(t_0)=0\}.
\end{equation*}
Then we have
\begin{equation*}
\int^1_0V(u)dt\rightarrow -\infty,\forall u_n\rightharpoonup u\in
\partial\Lambda.
\end{equation*}
\end{enumerate}
\end{lemma}
By Lemma 2.7 and 2.10, it's easy to prove:
\begin{lemma}\label{l2.8} Let $X$ be a  Banach space,let $F\subset X$ be a weakly closed subset. Suppose that $\Phi $
defined on $F$ is Gateaux-differentiable and  weakly lower
semi-continuous and bounded from below on $F$.
If $\Phi$ satisfies $(CPS)_{inf\Phi,F}$ condition or $(WCPS)_{inf\Phi,F}$ condition, and suppose that
\begin{equation*}
 \Phi(u_n)\rightarrow +\infty,u_n\rightharpoonup u\in \partial\Lambda,
\end{equation*}
 then $\Phi$ attains its infimum on $F$.
\end{lemma}
\begin{lemma}\label{l2.9} Let $X$ be a  Banach space, $F\subset X$ be a
 weakly closed subset. Suppose that $\phi(u)$
is defined on an open subset $\Lambda\subset X$ and is
Gateaux-differentiable on $\Lambda$ and weakly lower
semi-continuous and bounded from below on $\Lambda\bigcap F$, if $\phi$ is coercive, that is $\phi(x)\rightarrow +\infty$ as $||x||\rightarrow +\infty$, and
suppose that
\begin{equation*}
\phi(u_n)\rightarrow +\infty,u_n\rightharpoonup u\in \partial\Lambda,
\end{equation*}
 then $\phi$ attains its infimum on $\Lambda\bigcap F.$
\end{lemma}

\section{The Proof of Theorem 1.7}
\setcounter{section}{3} \setcounter{equation}{0}

Let
\begin{equation*}
\partial\Lambda_0=\{u\in H^1=W^{1,2}(R/Z,R^n),u(t+1/2)=-u(t),\exists t_0 , u(t_0)=0\}.
\end{equation*}
\begin{lemma}\label{l3.1} Assume $(A4)$ holds,
 then for any weakly convergent sequence $u_n\rightharpoonup u\in \partial\Lambda_{0}$, there holds
 \begin{equation*}
  f(u_n)\rightarrow +\infty.
 \end{equation*}
 \end{lemma}

 \begin{proof}
 Similar to the proof of Zhang[40].
 \end{proof}
\begin{lemma}\label{l3.2}
$F\bigcap\Lambda$ are weakly closed subset in $H^1$.
\end{lemma}
\begin{proof}
 Let $\{u_n\}\subset F\bigcap\Lambda$ be a weakly
convergent sequence, we use the embedding theorem to know which
uniformly converges to $u\in H^{1}$.

 Now we claim
$u\in\Lambda$, and then it's obviously that $u\in F$. In fact, if
$u\in\partial\Lambda$. By $V(q)\rightarrow -\infty,q\rightarrow 0$ and the condition $(A4)$ we have
\begin{equation*}
-V(u)\geq C_1|u|^{-\beta},0<|u|<r'<r.
\end{equation*}
So $V(u)$  satisfies Gordon's Strong Force Condition, by his Lemma, we have
\begin{equation*}
\int^1_0-V(u_n)dt\rightarrow +\infty,\forall u_n\rightharpoonup u\in
\partial\Lambda
\end{equation*}
Condition $(A4)$ implies
\begin{equation*}
V(u_n)+\frac{1}{2}<V^{\prime}(u_n),u_n>\geq
(1-\frac{\beta}{2})V(u_n).
\end{equation*}
Hence
\begin{equation*}
h=\int^1_0[V(u_n)+\frac{1}{2}<V^{\prime}(u_n),u_n>]dt\rightarrow
+\infty.
\end{equation*}
This is a contradiction.
\end{proof}

\begin{lemma}\label{l3.3}
$f(u)$ is weakly lower semi-continuous on $F\bigcap\Lambda_0$
\end{lemma}
\begin{proof}
 For any $u_n\subset F$: $u_n\rightharpoonup u$, then by
Sobolev's embedding Theorem, we have the uniformly convergence:
\begin{equation*}
|u_n(t)-u(t)|_{\infty}\rightarrow 0.
\end{equation*}
\begin{enumerate}
\item[(i).] If $u\in\Lambda_0$, then by $V\in C^1(R^n\backslash \{0\},R)$, we have
\begin{equation*}
|V(u_n(t))-V(u(t)|_{\infty}\rightarrow 0
\end{equation*}
And by the weakly lower semi-continuity for norm, we have
\begin{equation*}
\liminf \|u_n\|\geq\|u\|.
\end{equation*}
 Hence
\begin{equation*}
\liminf f(u_n)=liminf(\frac{1}{2}\int^1_0|\dot{u_n}|^2dt)\int^1_0(h-V(u_n))dt.
\end{equation*}
\begin{equation*}
\geq\frac{1}{2}\int^1_0|\dot{u}|^2dt\int^1_0(h-V(u))dt=f(u).
\end{equation*}
\item[(ii).] If $u\in\partial\Lambda_0$, then by $V$ satisfying Gordon's
Strong Force condition, we have
\begin{equation*}
\int^1_0-V(u_n)dt\rightarrow +\infty,\forall u_n\rightharpoonup u\in
\partial\Lambda_{0}.
\end{equation*}
  \begin{enumerate}
\item[(1).] if $u\equiv 0$, then
\begin{equation*}
|u_n|_{\infty}\rightarrow 0,n\rightarrow +\infty.
\end{equation*}
 Then similar to the proof in [40], we have
\begin{equation*}
f(u_n)\geq 6|{u_n}|_{\infty}^{2-\beta}\rightarrow +\infty,n\rightarrow
+\infty.
\end{equation*}
 So in this case we have
\begin{equation*}
 \liminf f(u_n)=+\infty\geq f(u).
\end{equation*}
\item[(2).] if $u\not=0$.
 By the weakly lower semi-continuity for norm, we have
 \begin{equation*}
 \liminf \|u_n\|\geq\|u\|>0 .
 \end{equation*}
 So by Gordon's Lemma, we have
\begin{eqnarray*}
\liminf f(u_n)&=&\liminf(\frac{1}{2}\int^1_0|\dot{u_n}|^2dt)\int^1_0(h-V(u_n))dt=+\infty\\
&=&\frac{1}{2}\int^1_0|\dot{u}|^2dt\int^1_0(h-V(u))dt=f(u).
\end{eqnarray*}
  \end{enumerate}
\end{enumerate}
\end{proof}

\begin{lemma}\label{l3.4} The functional $f(u)$ has
positive lower bound on $F$.
\end{lemma}
\begin{proof} By the definitions of $f(u)$ and $F$ and the
assumption $(A2)$, we have
\begin{equation*}
f(u)=\frac{1}{4}\int^1_0|\dot{u}|^2dt\int^1_0(V^{\prime}(u)u)dt\geq 0,\forall u\in F.
\end{equation*}
\end{proof}

By the definitions of the functional $f(u)$ and its domain $\Lambda_0$, and the conditions on the energy $h>0$ and the potential $V(u)<0$, it's easy to prove the following lemma.
\begin{lemma}\label{l3.5}
The functional $f(u)$ is coercive.
\end{lemma}

 Furthermore, we claims that
\begin{equation*}
c=\inf_{F\bigcap\Lambda_0} f(u)>0,
\end{equation*}
since otherwise, $u_0(t)=const$ attains the infimum 0, then by
the symmetry of $\Lambda_0$, we have $u_0(t)\equiv o$, which
contradicts the definition and $(A4)$.
 Now by  Lemmas 3.1-3.4  and Lemma 2.11, we know $f(u)$
attains the infimum on $F$, and we know that the minimizer is
nonconstant.

\section{The Proof of Theorem 1.8}
\setcounter{section}{4} \setcounter{equation}{0}

In order to prove the Cerami-Palais-Smale type condition and get non-constant periodic solution in non-symmetrical case, we need to add a topological condition,
 we know that there are winding numbers (degrees)in the planar case, so we define
 \begin{equation*}
 \Lambda_1=\{u\in\Lambda,deg(u)\not=0\}
 \end{equation*}
\begin{lemma}\label{l4.3}
If $u_n\rightharpoonup u\in\partial\Lambda_1$,then $f(u_n)\rightarrow +\infty.$
\end{lemma}
\begin{proof} By $V$ satisfying Gordon's Strong Force condition, we have
\begin{equation*}
\int^1_0-V(u_n)dt\rightarrow +\infty,\forall u_n\rightharpoonup u\in
\partial\Lambda_{1}.
\end{equation*}
\begin{enumerate}
\item[(1).] If $u\equiv 0$, then by Sobolev's embedding Theorem, we have
\begin{equation*}
|u_n|_{\infty}\rightarrow 0, \ \ n\rightarrow +\infty.
\end{equation*}
 Then by $deg(u_n)\not=0$,we have $c>0$ such that
 \begin{equation*}
  c|u_n|_{\infty}\leq ||\dot{u}_n||_{L^2}
 \end{equation*}
 and $||\dot{u}_n||_{L^2}$ is an equivalent norm of $W^{1,2}$
 and
\begin{equation*}
f(u_n)\geq c|{u_n}|_{\infty}^{2-\beta}\rightarrow +\infty,n\rightarrow
+\infty.
\end{equation*}
 So in this case, we have
 \begin{equation*}
 \liminf f(u_n)=+\infty\geq f(u).
 \end{equation*}
\item[(2).] If $u\not=0$.
 By the weakly lower semi-continuity for norm, we have
 \begin{equation*}
 \liminf \|u_n\|\geq\|u\|>0 .
 \end{equation*}
 So by Gordon's Lemma, we have
\begin{eqnarray*}
\liminf f(u_n)&=&\liminf(\frac{1}{2}\int^1_0|\dot{u_n}|^2dt)\int^1_0(h-V(u_n))dt=+\infty\\
&=&\frac{1}{2}\int^1_0|\dot{u}|^2dt\int^1_0(h-V(u))dt=f(u).
\end{eqnarray*}
\end{enumerate}
\end{proof}

\begin{lemma}\label{l4.1}
 Under the assumptions of Theorem1.8,
 \begin{equation*}
 f(u)=\frac{1}{2}\int^1_0|\dot{u}|^2dt\int^1_0(h-V(u))dt
 \end{equation*}
 satisfies $(CPS)^{+}$ condition on $\Lambda_1$, that is :
 If $\{u_n\}\subset \Lambda_1$ satisfies
\begin{equation}
f(u_n)\rightarrow c>0,\ \ \ \
(1+\|u_n\|)f^{\prime}(u_n)\rightarrow O.\label{4.1}
\end{equation}
Then $\{u_n\}$ has a strongly convergent subsequence in $\Lambda_1$.
\end{lemma}
\begin{proof}
 Since $f^{\prime}(u_n)$ make sense, we know
\begin{equation*}
\{u_n\}\subset\Lambda_1.
\end{equation*}
We claim $\int^1_0|\dot{u}_n|^2dt$ is bounded. In fact, by
$f(u_n)\rightarrow c$, we have
\begin{equation}
-\frac{1}{2}\|\dot{u}_n\|^2_{L^2}\cdot\int^1_0V(u_n)dt\rightarrow
c-\frac{h}{2}\|\dot{u}_n\|^2_{L^2}\label{4.2}
\end{equation}

By $(A3)'$ we have
\begin{eqnarray}
<f^{\prime}(u_n),u_n>&=&\|\dot{u}_n\|^2_{L^2}\cdot\int^1_0(h-V(u_n)-\frac{1}{2}<V^{\prime}(u_n),u_n>)dt\nonumber\\
&\geq&\|\dot{u}_n\|^2_{L^2}\int^1_0[h-\frac{\mu_2}{2}-(1-\frac{\alpha}{2})V(u_n)]dt\label{4.3}
\end{eqnarray}
By (4.2) and (4.3) we have
\begin{eqnarray}
<f^{\prime}(u_n),u_n>&\geq&(h-\frac{\mu_2}{2})\|\dot{u}_n\|^2_{L^2}+(1-\frac{\alpha}{2})(2c-h\|\dot{u}_n\|^2_{L^2})\nonumber\\
&=&(\frac{\alpha}{2}h-\frac{\mu_2}{2})\|\dot{u}_n\|^2_{L^2}+C_1\label{4.4}
\end{eqnarray}
where $C_1=2(1-\frac{\alpha}{2})c,\alpha>
2,h>\frac{\mu_2}{\alpha}.$ So $\|\dot{u}_n\|_{2}\leq C_2.$

Then we claim $|u_n(0)|$ is bounded.

 We notice that
\begin{eqnarray}
f^{\prime}(u_n)\cdot
(u_n-u_n(0))&=&\int^1_0|\dot{u}_n|^2dt\int^1_0(h-V(u_n))dt\nonumber\\
&&-\frac{1}{2}\int^1_0|\dot{u}_n|^2dt\int^1_0<V^{\prime}(u_n),u_n-u_n(0)>dt\nonumber \\
&=&2f(u_n)-\frac{1}{2}\int^1_0|\dot{u}_n|^2\int^1_0<V^{\prime}(u_n),u_n-u_n(0)>dt\label{3.11}
\end{eqnarray}
If $|u_n(0)|$ is unbounded,then there is a subsequence, still
denoted by $u_n$ s.t. $|u_n(0)|\rightarrow +\infty$. Since
\begin{equation*}
\|\dot{u}_n\|\leq M_1,
\end{equation*}
then
\begin{eqnarray}
\min_{0\leq t\leq
1}|u_n(t)|&\geq&|u_n(0)|-\|\dot{u}_n\|_2\rightarrow +\infty,
\rm{as}\ n\rightarrow +\infty\label{3.12}
\end{eqnarray}
 By Friedrics-Poincare's inequality and condition (P1) ,we have
\begin{eqnarray}
\int^1_0|\dot{u_n}(t)|^2dt&\geq&\pi^2\int^1_0|u_n(t)-u_n(0)|^2dt,\\
\int^1_0V^{\prime}(u_n)(u_n&-&u_n(0))dt\rightarrow 0,\\
f^{\prime}(u_n)\cdot (u_n&-&u_n(0))\rightarrow 0.
\end{eqnarray}

So $f(u_n)\rightarrow 0$, which contradicts $f(u_n)\rightarrow
c>0$, hence $u_n(0)$ is bounded, and
$\|u_n\|=\|\dot{u}_n\|_{L^2}+|u_n(0)|$ is bounded.
Furthermore,similar to the
proof of Ambrosetti-Coti Zelati([5]),$u_n$ strongly converges to
$u\in \Lambda$.
\end{proof}
It's easy to prove
\begin{lemma}\label{l4.2}
Under the assumption $(B1)'$, $f(u)\geq 0$ on
$\Lambda$, that is, $f$ has lower bound.
\end{lemma}

\begin{lemma}\label{l4.2}
Under the assumptions of Theorem1.8, $f(u)$ is weakly lower semi-continuous on the closure $\bar{\Lambda}$ of
$\Lambda$.
\end{lemma}
Now we can prove our Theorem 1.8, in fact, by Lemma 4.1, we know that the infimum of $f$ on $\Lambda_1$ is equal to the infimum of $f$ on the closure of $\Lambda_1$. Furthermore, we can prove the infimum of $f$ on $\Lambda_1$ is great than zero, otherwise if it is zero, the corresponding minimizer must be constant, then the winding number is zero, which is a contradiction. Now by the above Lemmas, especially Lemma 2.11, we know that $f$ attains the positive infimum on $\Lambda_1$ and the corresponding minimizer must be non-constant.

\section*{Acknowledgements section}

NSF of China and the Grant for the Advisors of Ph.D students.

\end{document}